\documentstyle[preprint,prb,aps]{revtex}

\begin{document}

\draft \tighten

\preprint{FIMAT-5/95}

\title{Physical nature of critical wave functions in Fibonacci systems}

\author{Enrique Maci\'{a}$^{*}$ and Francisco
Dom\'{\i}nguez-Adame$^{\dag}$}

\address{Departamento de F\'{\i}sica de Materiales,
Facultad de F\'{\i}sicas, Universidad Complutense,\\
E-28040 Madrid, Spain}

\maketitle

\begin{abstract}

We report on a new class of critical states in the energy spectrum of
general Fibonacci systems.  By introducing a transfer matrix
renormalization technique, we prove that the charge distribution of
these states spreads over the whole system, showing transport properties
characteristic of electronic extended states.  Our analytical method is
a first step to find out the link between the spatial structure of these
critical wave functions and the quasiperiodic order of the underlying
lattice.

\end{abstract}

\pacs{PACS numbers: 71.25.-s, 61.44.+p, 71.90.+q}

\narrowtext

The notion of {\em critical} wave function (CWF) has evolved
continuously since its introduction in the study of aperiodic systems
\cite{Ostlund}, leading to a somewhat confusing situation.  For
instance, references to self-similar, chaotic, quasiperiodic,
lattice-like or quasilocalized CWFs can be found in the literature
depending on the different criteria adopted to characterize them
\cite{KB,KST,Severin,Ryu,Gell}.  Generally speaking, CWFs exhibit a
rather involved oscillatory behavior, displaying strong spatial
fluctuations which show distinctive self-similar features in some
instances.  As a consequence, the notion of an envelope function, which
has been most fruitful in the study of both extended and localized
states, is mathematically ill-defined in the case of CWFs, and other
approaches are required to properly describe them and to understand
their structure.

Most interestingly, the possible existence of {\em extended} critical
states in several kinds of aperiodic systems, including both
quasiperiodic \cite{K&A,X&D,Kumar,Chak92} and non-quasiperiodic ones
\cite{Severin,Sil}, has been discussed in the last few years spurring
the interest on the precise nature of CWFs and their role in the physics
of aperiodic systems.  From a rigorous mathematical point of view the
nature of a state is uniquely determined by the {\em measure} of the
spectrum to which it belongs.  In this way, since it has been proven
that Fibonacci lattices have purely singular continuous energy spectra
\cite{Suto}, we must conclude that the associated electronic states
cannot be, strictly speaking, extended in the Bloch's sense.  This
result holds for other aperiodic lattices (Thue-Morse, period doubling)
as well \cite{Belli}, and it may be a general property of the spectra of
self-similar aperiodic systems \cite{Bovier}.  On the other side, from a
physical viewpoint, the states can be classified according to their {\em
transport properties} which, in turn, are determined by the spatial
distribution of the wave function amplitudes ({\em charge
distribution}).  Thus, conducting systems are described by periodic
Bloch states, whereas insulating systems are described by exponentially
decaying wave functions corresponding to localized states.  In this
sense, since the amplitudes of CWFs in a Fibonacci lattice do not tend
to zero at infinity but are bounded below throughout the
system\cite{Iochum}, one may expect their physical behavior to be more
similar to that corresponding to extended states than to localized ones.

In this letter we are going to show analytically that a subset of the
CWFs belonging to general Fibonacci systems are extended from a physical
point of view.  This result widens the notion of extended wave function
to include electronic states which {\em are not} Bloch functions, and it
is a relevant first step to clarify the precise manner in which the
quasiperiodic order of Fibonacci systems influences their transport
properties \cite{EXCITON}.  We shall begin the investigation of this
issue by introducing two novel approaches.  In the first place, we
present a new renormalization approach opening, in a natural way, an
algebraic formalism which allows us to give a detailed {\em analytical}
account of the transport properties of CWFs for certain particular
values of the energy.  In the second place, we relate the spatial
structure of CWFs to the topology of the underlying lattice by means of
the study of the {\em Fourier} spectrum of the wave function amplitudes
and multifractal techniques.

The formalism we are going to introduce is based on the transfer matrix
technique where the solution of the Schr\"odinger equation is obtained
by means of a product of $2 \times 2$ matrices.  Real-space
renormalization group approaches, based on decimation schemes, have
proved themselves very successful in order to numerically obtain the
energy spectrum of deterministic aperiodic systems \cite{N&N,PRBE}.  The
convenience for such procedures stems from the fact that, by
construction, a given transfer matrix relates only three consecutive
sites along the lattice, so that by decimating the original chain into
successively longer blocks we are able to describe the electronic state
corresponding to sites more and more farther apart.  In this context,
the key point of our procedure consists of the fact that we {\em
renormalize the set of transfer matrices instead of the lattice itself}.
Since these matrices contain all the relevant information concerning the
dynamics of the electrons, our approach becomes specially well suited to
describe the characteristic features associated to the long-range order
of the underlying Fibonacci system for, as we will see below, it
preserves the original quasiperiodic order of the lattice at any stage
of the renormalization process.

Let us start by considering a general Fibonacci system in which both
diagonal and off-diagonal terms are present in the Hamiltonian
\cite{PRBE,Sire}
$$
H=\sum_n \left\{ V_n |n \rangle \langle n| + t_{n,n+1} |n
\rangle \langle n+1| + t_{n,n-1} |n \rangle \langle n-1|\right\},
\label{Hamilton}
$$
where $V_n$ is the on-site energy and $t_{n,n\pm1}$ are the
nearest-neighbor hopping integrals. This Hamiltonian can be rewritten in
terms of the following matrices
$$
X \equiv \left(
\begin{array}{cc} {E-\beta\over t_{AB}} & -1 \\ 1 & 0 \end{array}
\right), \, \,
Y \equiv \left( \begin{array}{cc} \gamma^{-1} \,
{E-\alpha\over t_{AB}} & - \gamma^{-1} \\ 1 & 0
\end{array} \right),
$$
\begin{equation}
Z \equiv \left( \begin{array}{cc}
{E-\alpha\over t_{AB}} & -\gamma \\ 1 & 0 \end{array}\right), \, \,
W \equiv \left( \begin{array}{cc} {E-\alpha\over t_{AB}} & -1 \\ 1 & 0
\end{array} \right),
\label{matrix}
\end{equation}
where $E$ is the electron energy, $\alpha\ (\beta)$ denote the on-site
energies of sites A (B), $t_{AB}=t_{BA}$ and $t_{AA}$ are the
corresponding hopping integrals and $\gamma \equiv t_{AA}/ t_{AB}>0$.
Making use of these matrices we can {\em translate} the atomic sequence
ABAAB \ldots\ describing the topological order of the Fibonacci lattice
to the transfer matrix sequence XYZXWXYZXYZXW \ldots\ describing the
behavior of electrons moving through it.  In spite of its greater
apparent complexity, we realize that by renormalizing this TMS according
to the blocking scheme R$_A \equiv$ XYZ and R$_B \equiv$ XW, we get the
considerably simplified sequence R$_A$R$_B$R$_A$R$_A$R$_B$ \ldots\ The
subscripts in the Rs matrices are introduced to emphasize the fact that
the renormalized TMS is also arranged according to the Fibonacci
sequence and, consequently, the topological order present in the
original lattice is preserved by the renormalization process.  Let
$N=F_n$ be the number of lattice sites, where $F_n$ is a Fibonacci
number obtained from the recursive law $F_k=F_{k-1}+F_{k-2}$, with
$F_1=1$ and $F_0=1$.  It can then be readily shown that the renormalized
TMS contains n$_A \equiv F_{n-3}$ matrices R$_A$ and n$_B \equiv
F_{n-4}$ matrices R$_B$.

We will now use several properties of the Rs matrices to develop our
procedure.  Firstly, they are {\em unimodular} (i.\ e.\ their
determinant is one) for {\em any} choice of the system parameters and
for {\em any} value of the electron energy.  Secondly, they commute for
certain values of the energy.  In fact, after some algebra we get
\begin{equation}
[R_A,R_B] = \frac{P(E)}{\omega^2 \gamma} \,
\left( \begin{array}{cc} -1 & E+\alpha \\ 0 & 1
\end{array} \right),
\label{commutator}
\end{equation}
where we have defined the origin of energies in such a way that
$\beta=-\alpha$ and $t_{AB} \equiv 1$, and
\begin{equation}
P(E)=(\gamma^2-1)(E+\alpha)[(E-\alpha)^2-\omega^2]+
2\alpha \omega^2 \gamma^2 ,
\label{poly}
\end{equation}
with $\omega^2 \equiv (1+\gamma)t^2_{AB}$.  This commutator considerably
simplifies for the two cases mostly discussed in the literature, namely
the on-site ($\gamma \equiv 1$) and transfer ($\alpha \equiv 0$) models.
The expression (\ref{commutator}) shows that the on-site model is {\em
intrinsically} non-commutative, for the commutator vanishes only in the
trivial periodic case.  On the contrary, there exist three energy values
for which the R matrices commute in the transfer model, corresponding to
$E=0$ and $E=\pm \omega$.  Most interestingly, since $P(E)$ is a real
cubic polynomial in $E$, there exists {\em at least one} energy
satisfying the relation $P(E)=0$ for any realization of the mixed model
(i.\ e.\ for any $\gamma$ value).  For these energies the condition
$[R_A,R_B]=0$ is fulfilled and, making use of the Cayley-Hamilton
theorem for unimodular matrices \cite{theo} the global transfer matrix
of the system, $M(N) \equiv$ R$_A^{n_A}$R$_B^{n_B}$, can be explicitly
evaluated in terms of Chebyshev polynomials of the second kind,
$U_m(x)$, where $x$ is Tr(R)$/2$ .  Alternatively, the required power
matrices can be evaluated by diagonalizing them to a common basis.
{}From the knowledge of $M(N)$ the condition for the considered energy
value to be in the spectrum, $|$Tr$[M(N)]| \leq 2$, can be readily
checked and, afterwards, relevant parameters describing their transport
properties can be determined explicitly.  In this way, given any
arbitrary Fibonacci lattice, we are able to obtain a subset of its
energy spectrum whose eigenstates can be studied analytically.

Although our approach is completely general, the solutions of the
algebraic equation $P(E)=0$ are rather involved in most cases.  For
convenience, model parameters for which the algebra considerably
simplifies will be discussed henceforth.  In Fig.~\ref{fig1} we show the
charge distribution of two electronic states, for which the R matrices
commute, corresponding to Fibonacci lattices with $N=F_{16}=1597$.
Figure~\ref{fig1}(a) corresponds to the energy $E=2$ and lattice
parameters $\gamma=3$ and $\alpha=1$.  Figure~\ref{fig1}(b) shows the
result for the energy $E=\sqrt{1+ \gamma}$ for a transfer model ($
\alpha=0$) with $\gamma=1+\sqrt{2}$, the so-called silver mean.  The
overall periodic-like behavior of the wave function amplitudes, which we
have calculated exactly with the aid of our matrix formalism, clearly
indicates their extended character.

In order to discuss this point from a more rigorous physical perspective
we will focus our attention on the transmission coefficient, $\tau(E)$.
For the transfer model the global transfer matrices corresponding to the
energies $E=\pm \omega$ in the interval $0 \leq \gamma \leq 3$ can be
expressed, after lengthy algebra, in the closed form
\begin{equation}
M(N,\pm \omega)=R_A^{-p}=
\left( \begin{array}{cc} U_p(\theta^{\pm}) & U_{p-1}(\theta^{\pm})
\\ -U_{p-1}(\theta^{\pm})  & -U_{p-2}(\theta^{\pm})
\end{array} \right) ,
\label{e1}
\end{equation}
where $p \equiv 2n_B-n_A = F_{n-6}$ and $2 \cos{\theta^{\pm}}= \mp
\omega$.  From expression (\ref{e1}) we get Tr$[M(N)]=2 \cos{p
\theta^{\pm}}$ and, consequently, we can ensure that the energies $E=\pm
\omega$ belong to the spectrum in the quasiperiodic limit ($N
\rightarrow \infty$).  Now, we proceed to the calculation of the
transmission coefficient by embedding the Fibonacci lattice in an
infinite periodic arrangement of identical atoms connected by hopping
integrals $t$.  In this way we obtain
\begin{equation}
\tau(\pm \omega)=
\frac{1}{1+
\frac{4\omega^2(t\mp1)^2}{(4-\omega^2)(4t^2-\omega^2)}
\sin^2{p \theta^{\pm}}}.
\end{equation}
Two important conclusions can be drawn from this expression.  In the
first place, the transmission coefficient is always bounded below for
{\em any} lattice length, which proves the true extended character of
the related eigenstates.  In the second place, we observe that the
transparency condition $\tau=1$ is obtained, for any value of $t$, for
certain chain sizes satisfying $p \theta^{\pm}=k \pi$, $k=1,2 \ldots$,
which in turn implies $\pm \omega=\mp 2\cos (k \pi/p)$.  In this way the
transparent states $\tau=1$ can be classified according to a well
defined scheme determined by the integers $k$ and $p$.  Thus the state
shown in Fig.~\ref{fig1}(b) corresponds to the choice $k=1$ and $p=8$.

After having discussed the transport properties of this class of
critical wave functions, we turn our attention to their spatial
structure.  To this end we shall consider the example shown in
Fig.~\ref{fig2} which corresponds to the energy $E=\sqrt{2 \sigma}$,
where $\sigma \equiv (\sqrt{5}+1)/2$ is the so-called golden mean, and
model parameters $\gamma=\sqrt{5}$ and $\alpha=0$.  In
Fig.~\ref{fig2}(a) we show the overall charge distribution through a
lattice with $N=F_{21}=17\,711$ sites.  From this plot we notice the
existence of two different superimposed structures.  In fact, a
periodic-like long-range oscillation with a typical wavelength of about
$18\,000$ sites is observed to modulate a quasiperiodic series of
short-range minor fluctuations of the wave function amplitude, typically
spreading over $122$ lattice sites.  This qualitative description
receives a quantitative support from the study of its Fourier transform,
as it is shown in Fig.~\ref{fig2}(b).  In fact, we observe two major
components in the Fourier spectrum corresponding to the low and high
frequency regions, respectively.  In the low frequency region two
relevant features at frequencies $\nu_1\simeq 1.11\times 10^{-4}$ and
$\nu_2\simeq 8.18\times 10^{-3}$ are present, in agreement with the
short- and long-range structures of the charge distribution observed in
Fig.~\ref{fig2}(a).  On the other side, in the high frequency region we
observe a series of features grouped around frequency positions given
by successive powers of the inverse golden mean $\sigma^{-1} \equiv
(\sqrt{5}-1)/2$.  These features are labelled correspondingly in
Fig.~\ref{fig2}(b) and are characteristic of the quasiperiodic nature of
the charge distribution of the considered states.  By this we mean that
we have observed that their weights in the Fourier spectrum increase and
their positions approach the reported powers of $\sigma^{-1}$ as the
system size increases.

Finally, to gain further insight into the behavior of the wave function
at all length scales we have performed a multifractal analysis of the
states belonging to the subset $E= \pm \omega$.  The amplitude
distribution of the electronic states has been characterized by the
scaling of moments $\mu_q(N)$ of order $q$, associated to their charge
distribution, with the system size (for a definition of those moments
see, e.g, Ref.~\onlinecite{Angel}).  The multifractal dimension $D_q$ is
determined via the scaling $\mu_q(N)\sim N^{(1-q)D_q}$ for $q\neq 0$.
In all cases studied we have found that $D_q=1$, for all $q$, and for
system sizes as large as $N=F_{30}=1\,346\,269$.  Thus the {\em lack of
multifractality} along with the fact that $D_q$ equals the spatial
dimension proves our claim that these states uniformly spread over the
whole system.

In summary, in this letter we prove the existence of a subset of the
singular continuous spectrum characteristic of Fibonacci systems whose
eigenstates are extended in the physical sense previously discussed.
This we have shown by means of a transfer matrix renormalization
technique which allows us to unveil the effects of short-range
correlations by grouping ABA sites and AB sites into the matrices R$_A$
and R$_B$, respectively.  In this sense we can properly state that these
states are characteristic of the quasiperiodic order of the underlying
lattice.  Interestingly we note that similar results concerning extended
states in Thue-Morse chains have been recently reported in the
literature \cite{Chakra}.  We wish to stress that the algebraic approach
presented in this work can be extended in a straightforward manner to
other kinds of aperiodic systems based on substitution sequences, and
therefore it can be relevant in order to attain a unified treatment of
physical properties of aperiodic systems.  In closing, we note that the
transparency condition $\tau=1$ obtained for extended Fibonacci states
is similar to that appearing in random dimer models \cite{RDM} and this
fact suggests that sets of extended states may arise in general
aperiodic systems.

\acknowledgments

It is with great pleasure that we thank Angel S\'anchez for far reaching
suggestions and illuminating conversations on the role of correlation in
the physics of aperiodic systems.  We also thank M.\ V.\ Hern\'andez for
interesting discussions. This work is supported by CICYT under project
No.\ MAT95-0325.

%REFERENCES

%FIGURES

\begin{figure}
\caption{Electronic charge distribution in Fibonacci lattices with
$N=F_{16}$ and (a) $\gamma=3$, $\alpha=1$, $E=\omega=2$ and (b)
$\gamma=1+\protect{\sqrt{2}}$, $\alpha=0$, $E=\omega=
\protect{\sqrt{2+\protect{\sqrt{2}}}}$.
Insets show finer details of the squared wave functions.}
\label{fig1}
\end{figure}

\begin{figure}
\caption{(a) Electronic charge distribution in a Fibonacci lattice with
$N=F_{21}$, $\gamma=\protect{\sqrt{5}}$, $\alpha=0$, and $E=\omega =
\protect{\sqrt{1+\protect{\sqrt{5}}}}$.  Inset shows finer details of
the squared wave function.  (b) The corresponding Fourier spectrum.
Inset shows the lower frequency region.}
\label{fig2}
\end{figure}

\end{document}